# Electronic Structure of Lanthanide Scandates

Christopher A. Mizzi, Pratik Koirala, Laurence D. Marks*

Department of Materials Science and Engineering, Northwestern University, Evanston IL, 60208, USA

## Abstract

X-ray photoelectron spectroscopy, ultraviolet photoelectron spectroscopy and density functional theory calculations were used to study the electronic structure of three lanthanide scandates: $GdScO_3$, $TbScO_3$, and $DyScO_3$. X-ray photoelectron spectra simulated from first principles calculations using a combination of on-site hybrid and GGA+U methods were found to be in good agreement with experimental x-ray photoelectron spectra. The hybrid method was used to model the ground state electronic structure and the GGA+U method accounted for the shift of valence state energies due to photoelectron emission via a Slater-Janak transition state approach. From these results, the lanthanide scandate valence bands were determined to be composed of Ln4f, O2p, and Sc3d states, in agreement with previous work. However, contrary to previous work the minority Ln4f states were found to be located closer to, and in some cases at, the valence band maximum. This suggests that minority Ln4f electrons may play a larger role in lanthanide scandate properties than previously thought.

*Corresponding Author, Email: L-marks@northwestern.edu

# 1 Introduction

Lanthanide scandates possess distorted orthorhombic perovskite structures with space group *Pbmn* at room temperature and atmospheric pressure [1]. They exhibit low-temperature magnetic ordering [2], a highly anisotropic magnetic response [3] and polar phonons [4] among other interesting properties. Their wide use as thin film substrates has led to discoveries such as an enhanced ferroelectric response in epitaxially grown $SrTiO_3$ [5], and lanthanide scandates have been explored as potential gate oxide materials due to their high dielectric constants [6,7].

To understand these and other electronic and magnetic properties of lanthanide scandates, it is necessary to theoretically and experimentally characterize their electronic structures, particularly their valence bands. Density functional theory (DFT) has been successfully used for modeling electronic structure in many oxides, but has had varying levels of success in modeling correlated electron systems (such as the 4f electrons in lanthanide scandates) due to the difficulty of describing the correlated behavior in a mean-field framework. Experimental photoelectron spectroscopy techniques, such as x-ray photoelectron spectroscopy (XPS) and ultraviolet photoelectron spectroscopy (UPS) are well-suited to examine the electronic structure of correlated electron materials because emitted photoelectrons provide a means to probe the occupied density of states (DOS) [8,9]. Ideally, photoelectron spectroscopy can validate (or invalidate) DFT methods.

Previous studies have reported experimentally measured XPS valence band spectra of lanthanide scandates [2,10-12] and simulations of such spectra from first principles calculations [2] using generalized gradient approximation + Hubbard U methods (GGA+U). The prevailing interpretation is that the valence band is comprised of lanthanide 4f (Ln4f), scandium 3d (Sc3d), and oxygen 2p (O2p) states, although there is some ambiguity regarding the positions of these states stemming from the inherent difficulty of treating 4f states in DFT. For example, DFT calculations that include Hubbard U terms have led to the placement of the minority 4f states in the band gap or to very localized 4f states [2] well below the O2p states. Consequently, 4f states in lanthanide scandates have often been treated as atomic-like states far from the valence band maximum (VBM), which is taken to be dominated by O2p states.

Here, we present experimental electronic structure data acquired with XPS and UPS for $GdScO_3$, $TbScO_3$, and $DyScO_3$. We have also simulated the XPS spectra in all three of these materials based upon first principles calculations using an on-site hybrid method coupled with GGA+U. The on-site hybrid approach gives a reasonable description of the ground state electronic structure. The GGA+U method is used in an unconventional but rigorous way to model the effect of the photoelectron hole, in what is essentially a Slater-Janak transition state method [13-16]. We found reasonably good agreement between our experimental and simulated spectra and determined that the general features of the valence band—i.e. a valence band composed of Ln4f, O2p, and Sc3d states—agree with previous work [2,10-12,17]. We also found that the UPS spectra support the interpretation that O2p states are delocalized throughout the valence band. However, our results indicated that the minority Ln4f states are located closer to, and in some cases at, the VBM suggesting that minority Ln4f electrons may play a larger role in lanthanide scandate properties than previously thought.

## 2 Methods

### 2.1 Sample Preparation

Commercially available 10 mm × 10 mm × 0.5 mm single crystalline substrates of [110] oriented $LnScO_3$ (Ln = Gd, Tb, Dy) were purchased from MTI Corp. These substrates were annealed at 1050°C for 10 hours in air to promote surface ordering. Following the 1050°C anneal, the color of the DyScO$_3$ changed from yellow to brown, whereas the other lanthanide scandates did not change color (GdScO$_3$ and TbScO$_3$ remained white). This color change is likely due to the presence of small concentrations of tetravalent Dy, and/or Dy and O vacancies [18,19]. Prior to XPS and UPS measurements, all samples were baked at 600°C for 6 hours in air to minimize surface contamination, then placed in the high vacuum chamber of the XPS system to degas overnight. No color change was observed following the 600°C bake.

## 2.2 X-ray and Ultraviolet Photoelectron Spectroscopy

XPS measurements were performed on an ESCALAB 250Xi equipped with a monochromated, micro-focused Al K-Alpha (1486.6 eV) x-ray source. A 180° double focusing hemispherical analyzer with a dual detector system was used in constant analyzer energy mode. In addition to general survey spectra to check for impurities, higher resolution XPS spectra were acquired with a pass energy of 20 eV, step size of 0.1 eV, spot size of 650 μm, and averaged over 10-20 scans. UPS spectra were taken on the same instrument using a He II (40.8 eV) UV source with a photon flux of ~ 1.5 x 10$^{12}$ photons/second, pass energy of 2 eV, step size of 0.05 eV, spot size of ~ 1.5 mm, and averaged over 20 scans.

Lanthanide scandates were found to charge significantly under x-ray illumination, leading to a ~ 700 eV binding energy offset at room temperature. Therefore, XPS spectra were collected with an argon flood gun to minimize charging. The flood gun was operated at a beam voltage of 2 V, emission current of 50 μA, focus voltage of 20 V, and extractor bias of 30 V at a chamber pressure of ~ 10$^{-7}$ mbar. Small amounts of charging persisted even when a flood gun was used, so spectra were shifted using the adventitious C1s peak centered at 285 eV when this peak was sufficiently intense. Carbon quantification was performed for the sample with highest carbon coverage, GdScO$_3$, and it was determined that there was approximately a monolayer of carbon on that sample. In cases where the adventitious carbon intensity was too low (TbScO$_3$ and DyScO$_3$), the spectra were shifted by centering the Sc3p peak at 30.8 eV as this peak was found to shift by the same energy as the C1s peak upon charging.

Charge compensation using the flood gun was not possible for the UPS measurements because artifacts from its use masked the relatively low photoelectron intensity. Correction of the energy scale by other means was also not possible because the incident photon energy was insufficient to cause C1s photoemission and the Sc3p peak was masked by secondary electron emission. As a result, only relative binding energies have meaning in the UPS spectra. It is also worth noting that DyScO$_3$ charged less than GdScO$_3$, which charged less than TbScO$_3$ which caused varying degrees of charge shifting in the UPS spectra.

## 2.3  Density Functional Theory Calculations

DFT calculations were performed with the all-electron augmented plane wave + local orbitals WIEN2K code [20]. Muffin tin radii of 1.68, 1.82, and 2.02 were used for O, Sc, and the lanthanides Gd/Tb/Dy, respectively, to minimize the inclusion of O2p tails within the metal muffin tins which perturbs the calculation of the exact-exchange corrections inside the muffin tins for Sc and Gd/Tb/Dy. The plane-wave expansion parameter RKMAX was 9.0. The electron density and (when appropriate) all atomic positions were simultaneously converged using a quasi-Newton algorithm [21]. Both an on-site hybrid approach [22,23] as well as a GGA+U approach using the PBEsol [24] functional were used, as described in more detail in the Results section.

## 3  Results

## 3.1  Experimental Photoelectron Spectroscopy

Figure 1 shows XPS spectra of GdScO$_3$, TbScO$_3$, and DyScO$_3$ after correction of the origin as described above. The 95% probe depth of photoelectrons contributing to these spectra is ~ 10 nm

(estimated as 3 times the inelastic mean free path [25]). In all three spectra, the peak at 30.8 eV corresponds to Sc3p states, the peaks near 20 and 28 eV are the Ln5p doublet, and the peak at 23 eV is the O2s peak. While the Sc3p peak location appears to be invariant for different lanthanide scandates, the Ln5p and O2s peaks increase in binding energy as the atomic number of the lanthanide species increases. This systematic increase in Ln5p and O2s binding energy is expected because the nuclear charge of the Ln species in $LnScO_3$ increases from Gd to Tb to Dy [8,9]. These peak assignments agree with existing literature [10-12].

The measured valence band XPS spectra (considered to be all states with a binding energy less than 15 eV) differ qualitatively between the three scandates. There are two major peaks at 4 eV and 8 eV in the $GdScO_3$ valence band spectrum. The $TbScO_3$ valence band spectrum consists of a sharp peak at 2 eV and a wide feature centered at 9 eV, and the $DyScO_3$ valence band spectrum has three major peaks at 4 eV, 6 eV, and 9 eV. Since the Ln4f cross-sections are orders of magnitude larger than the cross-sections of other states contributing to the valence band spectra at an incident x-ray energy of 1486.6 eV [26], these differences likely originate from the correlated behavior and rich multiplet structure of Ln4f states [27].

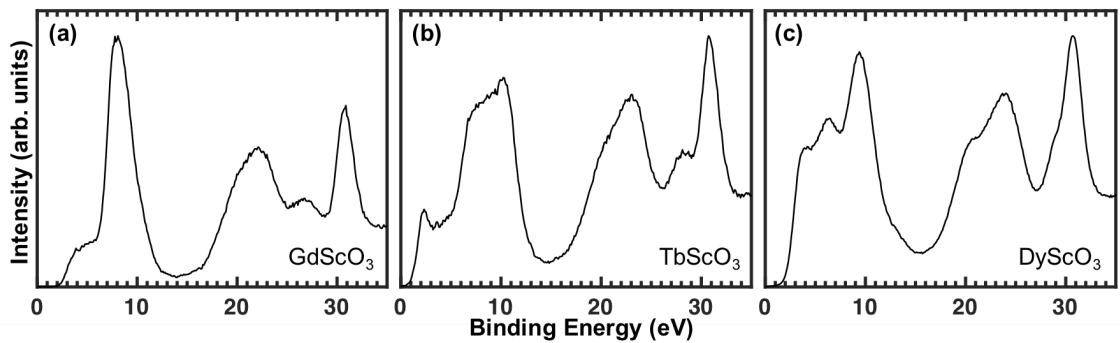

FIG. 1. Experimental x-ray photoelectron spectra acquired with 1486.6 eV incident x-rays for (a) $GdScO_3$, (b) $TbScO_3$, and (c) $DyScO_3$ after origin correction. Each valence band spectrum extends from 0 – 15 eV and is qualitatively different for the three lanthanide scandates: it consists of two major features at 4 eV and 8 eV in $GdScO_3$, two major features at 2 eV and 9 eV in $TbScO_3$, and three major features at 4 eV, 6 eV, and 9 eV in $DyScO_3$. The features from 15 eV – 35 eV correspond to the Ln5p doublet (20 eV and 28 eV), O2s (23 eV), and Sc3p peaks (30.8 eV).

To investigate this further, UPS spectra were acquired because O2p cross-sections are ~ 2 – 5 times larger than Ln4f cross-sections at the incident photon energies used for UPS [26]. Figure 2 shows UPS valence band spectra for GdScO$_3$, TbScO$_3$, and DyScO$_3$. The 95% probe depth of photoelectrons contributing to these spectra is ~ 1.2 nm [25], making this technique more surface sensitive than XPS. As mentioned in the Methods section, only relative binding energies have meaning in the data in Fig. 2 due to charging. Unlike the XPS spectra in Fig. 1 which showed qualitative differences across the three scandates, the UPS spectra are very similar: they all consist of two major peaks separated by 4 – 5 eV. Note that only energies below the onset of secondary electron emission are considered in Fig. 2.

Based upon the literature cross-sections, the XPS spectra suggest that the valence band has contributions from the Ln4f states and the UPS spectra indicate valence band contributions from O2p states. However, there is some ambiguity regarding the combination and location of these states. Additionally, the role of Sc3d states is unclear because the cross-sections of Sc3d states are too small relative to Ln4f and O2p at the incident energies used for these experiments. Therefore, DFT calculations were used to simulate the XPS spectra.

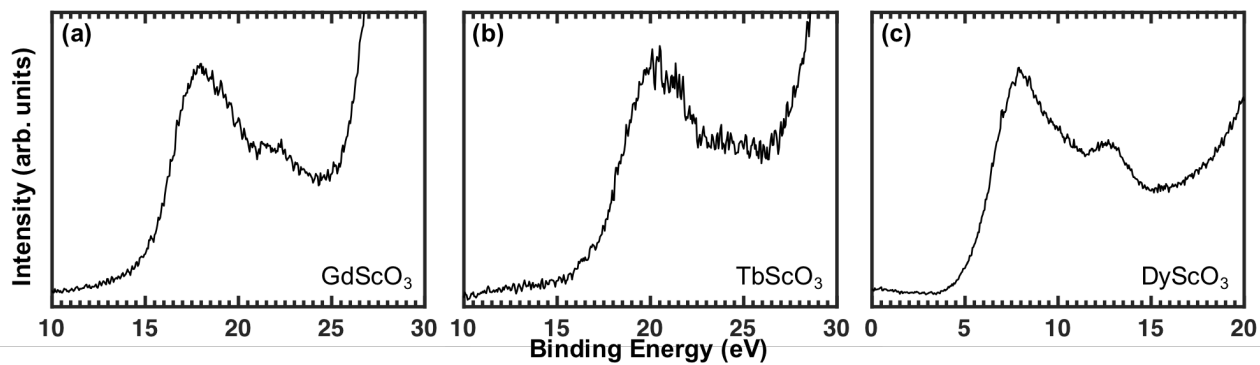

FIG. 2. Experimental ultraviolet photoelectron spectra acquired with 40.8 eV photons for (a) GdScO$_3$, (b) TbScO$_3$, and (c) DyScO$_3$. Because of charging, only relative binding energies have meaning in these spectra. The spectrum for each lanthanide scandate is qualitatively similar, consisting of two peaks separated by 4 – 5 eV.

**3.2    Photoelectron Spectroscopy Simulations**

A four-step method was utilized to model the XPS spectra. The general process is outlined in this paragraph with details on each step provided in subsequent paragraphs. First, the ground state electronic structure was determined using an on-site hybrid method. Second, the partial DOS (pDOS) calculated with an on-site hybrid method was modified to include the physics of the XPS process, i.e. relaxation effects arising from the removal of a photoelectron from an insulator [8,9], using GGA+U calculations. The reason why *both* hybrid and Hubbard U methods were used is explained in more detail below. Third, the intensities of this modified pDOS were scaled to account for the different photoionization cross-sections of each state. Since there is extensive evidence that literature cross-sections [26] are not reliable for the valence band, e.g. [28], literature cross-sections were used as starting values for pDOS scaling factors which were then varied to fit simulated spectra to experimental spectra. Fourth, the effects of instrument resolution, thermal broadening, and state lifetime were approximated by Gaussian broadening of the pDOS.

First, it is necessary to obtain an accurate representation of the ground state electronic structure. Conventional local density approximation + Hubbard U (LDA+U) or GGA+U methods are plausible methods for modelling 4f systems; however, the use of a Hubbard U value sufficient to obtain a reasonable band gap resulted in highly-confined, essentially atomic 4f occupied states which disagrees with our experimental data. This was true independent of whether a PBE [29] or mBJ [30] functional was used with a Hubbard U. Improved results were obtained using an on-site hybrid method, which uses an exact-exchange hybrid correction within the muffin tins.

A subtle question of any hybrid approach is the on-site hybrid fraction to use. It is now well established that the hybrid fraction is not a universal factor, but should vary [22,31]. To determine what values to use, the atomic positions and bulk optimized lattice constants were calculated using the on-site hybrid method with the PBEsol functional, and the hybrid fraction was varied to minimize the forces on the atoms using the known bulk positions [1]. All calculations were performed with a ferromagnetic ordering; the energy difference between ferromagnetic and anti-ferromagnetic orderings was minimal, as is expected since this is a weak energy term in lanthanide scandates. Optimized values of the on-site hybrid fractions were found to be 0.80 for Sc3d, 0.50 for Dy5d, and 0.375 for Dy4f states, within an uncertainty of approximately 0.05. The same values were used for the Ln and Sc species in TbScO$_3$ and GdScO$_3$. We note that a larger Dy4f on-site correction value when spin-orbit coupling was included led to a change in the spin-state which would disagree with the known magnetic moment for Dy in DyScO$_3$ [2].

Figure 3 shows pDOS from the on-site hybrid calculations where the VBM have been set to 0 eV and negative energies correspond to occupied states. The O2p and Sc3d contributions to the valence band are largely unchanged across all three lanthanide scandates: O2p states dominate and are delocalized, while the Sc3d states have a low pDOS throughout the valence band. On the contrary, the Ln4f pDOS are noticeably different for each material. Gd only possesses a single 4f peak since there are no minority 4f electrons present in trivalent Gd. The presence of minority 4f electrons leads to splitting of the majority 4f state and the appearance of a minority 4f state close to or at the VBM, as seen in the DyScO$_3$ and TbScO$_3$ pDOS. Other states were found to have negligible pDOS in the valence band and are excluded from this analysis.

From the on-site hybrid pDOS shown in Fig. 3, the band gaps were found to be 5.2 eV, 4.9 eV, and 5.3 eV for GdScO$_3$, TbScO$_3$, and DyScO$_3$, respectively. These values are similar to reported band gaps ([17] and references therein). The Sc3d and O2p contributions to the conduction band appear to be comparable for the three lanthanide scandates, consisting of high Sc3d and low O2p pDOS. As is expected, the unoccupied Ln4f states are all minority spin. It is worth noting that the GdScO$_3$ on-site hybrid calculations placed unoccupied Gd4f states in the band gap; however, this same behavior was not seen in TbScO$_3$ or DyScO$_3$. We suspect this is due to the system-dependent nature of the hybrid fraction (as is discussed in greater detail in the Discussion section). Ultimately, this does not impact the results shown here because only the occupied states from the on-site hybrid calculations are used for XPS simulations.

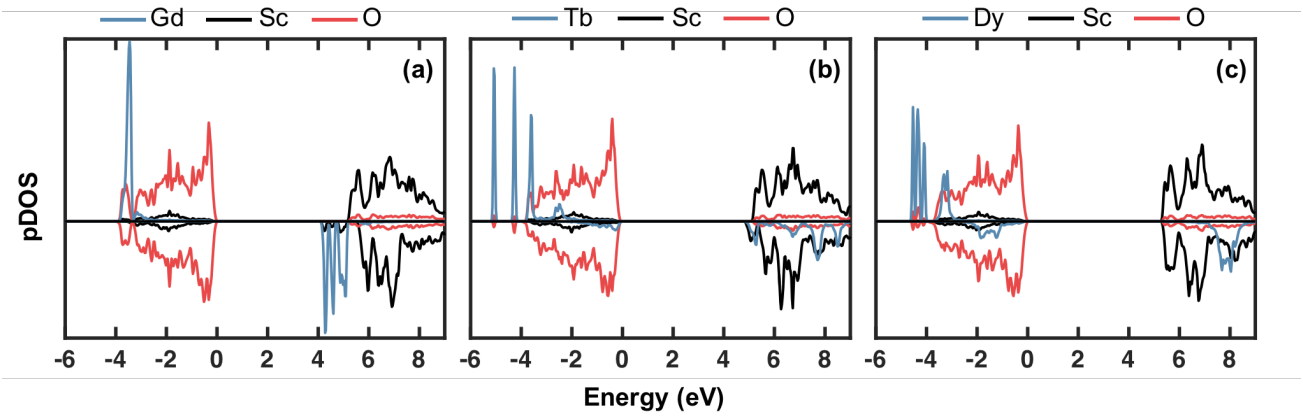

FIG. 3. (Color online) Partial density of states (pDOS) of (a) GdScO$_3$, (b) TbScO$_3$, and (c) DyScO$_3$ from on-site hybrid density functional theory calculations. The upper and lower panels correspond to the spin up and spin down pDOS, respectively. Blue lines indicate Ln4f, black lines correspond to Sc3d, and red lines show O2p pDOS. Negative energies are taken to be occupied states and the VBM has been set to 0 eV. For visual clarity, in this figure the O2p and Sc3d pDOS have been scaled by a factor of 10 and the pDOS of one symmetrically inequivalent oxygen is not shown.

Now relaxation effects are considered in these simulations. In photoelectron spectroscopies, the occupancy of a given state is altered by the removal of a photoelectron. It is well established that the resulting hole causes an energy level shift [8,9]. For core states this is typically accounted for by calculating the energy levels with a partial core hole, where for instance half a hole corresponds to the Slater approach [16]. Unfortunately, it is not easy in conventional DFT calculations to include half a hole in specific states, as it will migrate to the top of the valence band. There is a different approach which we used here which is effectively to use a Slater-Janak transition state method [13-16].

The Hubbard U approach adds a term which is rigorously defined as a correction such that the energy of the states is independent of occupancy [32]. If such a correction is calculated for specific valence states (e.g. 4f), this is equivalent to correcting for a valence band hole in that specific state. Therefore, the U term was calculated *ab initio* by placing the 4f electrons into the core and varying their occupancy. These values (U = 8.2 eV, J = 0 eV) were then used in GGA+U with PBEsol calculations which led to 4f energies with better agreement to our XPS experiments.

The 4f state positions from these GGA+U with PBEsol calculations (which accounted for the loss of a photoelectron during the XPS process) were then used to linearly scale the energies of the on-site hybrid pDOS (which provided a better representation of the ground state electronic structure) so that the on-site hybrid 4f state positions were at the same energy as the GGA+U 4f state positions. Specifically, linear scaling was performed using the ratio of the occupied Gd4f location calculated using the GGA+U and on-site hybrid methods. Gd was chosen to determine the linear scaling factor because it lacks the additional complexity associated with minority 4f electrons. This resulted in a linear scaling ratio of 1.7, which was also used for $TbScO_3$ and $DyScO_3$ simulations.

In order to compare these simulations with experimental spectra, differences in the photoionization cross-sections of different states must be considered. As mentioned before, there is pragmatic evidence that literature photoionization cross-sections are not accurate for valence bands [28]. Consequently, it is no surprise that using literature cross-sections (calculated with isolated atom *ab initio* methods [26], which neglect the effects of reconfiguration [8,9] and dielectric screening [33]), produced simulated spectra overly-weighted by Ln4f contributions when compared to our experimental spectra. Using the literature cross-sections as starting intensity scaling factors for the pDOS, scaling factor values were varied to find the best agreement with our experimental data. GdScO$_3$ was used for this fit as it has no minority 4f electrons, so the states close to the VBM can only be due to O2p states. From fitting to the GdScO$_3$ it was found that our simulated intensities matched experimental intensities if the O2p cross-section was increased by a factor of 10 and the Gd4f and Sc3d cross-sections were left unchanged. Based upon this GdScO$_3$ fit, the O2p cross-sections in the TbScO$_3$ and DyScO$_3$ pDOS were also increased by a factor of 10 relative to the tabulated cross-sections and the Ln4f and Sc3d cross-sections were left unchanged.

The effects of instrument resolution, thermal broadening, and state lifetime were approximated by applying a Gaussian broadening to the pDOS. Although a mixture of Lorentzian and Gaussian broadening is typically used [8,9], the inclusion of a Lorentzian broadening component had a negligible impact on the simulations. Gaussian broadening values were varied to find an optimal match to experimental spectra. For the aforementioned reasons, GdScO$_3$ was used to determine the O2p Gaussian broadening value of 0.30 eV. Gaussian broadening values of 0.57 eV (Gd4f), 0.30 eV (Tb4f), and 0.41 eV (Dy4f) were found to mimic experimental peak width.

Figure 4 shows the end-results of the four-step simulation method discussed above for GdScO$_3$, TbScO$_3$, and DyScO$_3$. All VBM have been set to 0 eV and each spectrum has been normalized by its maximum valence band intensity. Although there are discrepancies in the peak positions of ~ 0.5 – 1 eV between the experimental and simulated spectra shown in Fig. 4, these simulations provide a reasonable match with the experimental spectra. In particular, the GdScO$_3$ simulation reproduces a two peak structure, the TbScO$_3$ simulation exhibits a sharp peak at the VBM and a wide peak spanning ~ 5 eV, and the DyScO$_3$ simulation possesses a three peak structure.

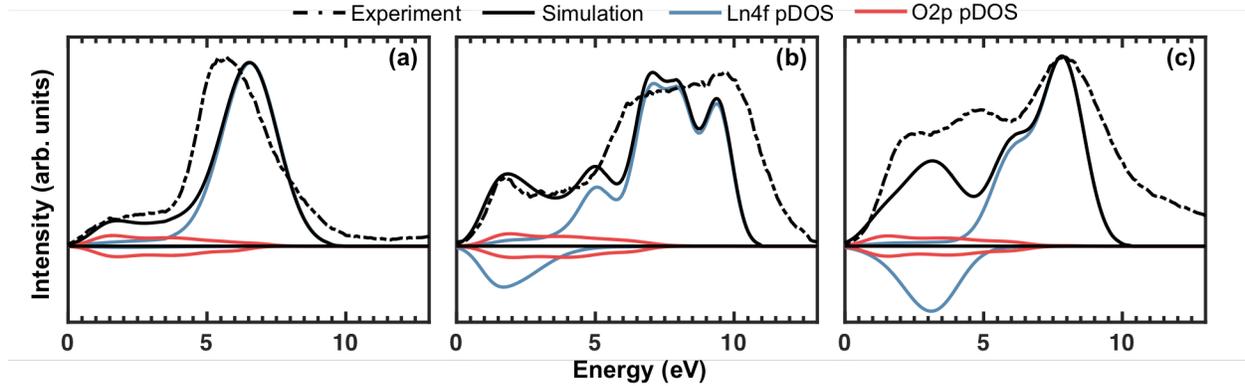

FIG. 4. (Color online) Comparison between experimental x-ray photoelectron spectra (dashed lines) and simulated x-ray photoelectron spectra (solid lines, black) for (a) GdScO$_3$, (b) TbScO$_3$, and (c) DyScO$_3$. The upper and lower panels correspond to the spin up and spin down pDOS, respectively. The simulations correspond to the sum of the O2p pDOS (solid lines, red), Sc3d pDOS (not shown), and Ln4f pDOS (solid lines, blue) after the simulation method described in the text. Each valence band maximum has been set to 0 eV and each spectrum has been normalized by its maximum valence band intensity. The simulations qualitatively agree with the experimental spectra and capture the main valence band features. Sc3d pDOS are not shown because Sc3d has negligible pDOS in the valence band.

## 4   Discussion

Previous experimental studies have placed Ln4f states at ~ 10 eV, O2p and Sc3d states at ~ 5 eV, and O2p states at the VBM [10-12]. Other studies used a combination of x-ray emission spectroscopy, x-ray absorption spectroscopy, XPS, and GGA+U calculations to conclude that the Ln4f states contribute to 10 eV and 5 eV features (depending on 4f occupancy), and Sc3d states and O2p states are delocalized from 5 eV to the VBM [2,17].

The experimental UPS spectra in Fig. 2 and the simulated O2p pDOS in Fig. 3 support the interpretation that the O2p states are delocalized throughout the valence band. Unfortunately, we were unable to experimentally probe the Sc3d states because the cross-section of these states is too low at the available incident XPS and UPS energies. Since the simulations seemed to accurately capture experimental behavior (i.e. 4f peak structure and O2p delocalization in all three scandates), it is reasonable to conclude that the Sc3d states have only a very small, delocalized contribution throughout the valence band from the pDOS.

The Ln4f states exhibit more complex behavior than the O2p or Sc3d states. It is common in the existing literature to treat Ln4f states as spectators: highly localized, essentially atomic states that exist at energies well below the VBM. The results presented in this paper indicate that the minority Ln4f and O2p states are at comparable energies close to or at the VBM, which means minority Ln4f electrons may play a larger role in lanthanide scandate properties. Moreover, the energy overlap between O2p and minority Ln4f states in the valence band indicates the possibility of substantial hybridization. The location of the minority Ln4f states with respect to the VBM also suggests that $TbScO_3$ is better described as a Mott-Hubbard insulator than $GdScO_3$ which is more definitively a charge-transfer insulator (according to the framework developed by Zaanen, Sawatzky, and Allen for transition metal compounds [34]).

It is worth noting that the quality of the match in Fig. 4 seemingly decreases with the addition of minority 4f electrons. It is likely that the deviations in $TbScO_3$ and $DyScO_3$ originate from multiplet splitting. However, given that this method approximated the XPS process, an inherently $N - 1$ electron process, with an N electron calculation and ignored multiplet effects, we argue that the simulations are in reasonable agreement with experimental spectra, certainly significantly better than in prior publications for these materials. Additionally, as a consequence of our simulation methods, O2p cross-sections were found to be an order of magnitude too small (or the Ln4f cross-sections were found to be an order of magnitude too large) when compared to experimental results. As discussed in the Results section, this is not surprising because the documented cross-sections correspond to atomic states and do not account for relaxation, hybridization, or dielectric screening.

For completeness, we also tested using a full hybrid approach using YSE0 based upon either the PBE [29] or PBEsol [24] functionals with a hybrid fraction in the range 0.25 – 0.30. While these calculations produced band gaps consistent with experimental measurements (better than the on-site approach), they resulted in DOS with minority Tb4f and Dy4f states bracketing the band gap, which did not agree as well with experimental data as the on-site approach. It is established that the fraction of exact-exchange that yields the best results is system dependent [35] and we suspect that the appropriate hybrid fraction for 4f electrons is probably higher because they are less accurately treated within the GGA component compared to more slowly varying s and p electrons.

In principle, a Rashba spin-orbit coupling term [36] should be included for the Ln4f states near the surface, which would probably lead to a spiral spin wave going in from the surface. Tests adding collinear spin-orbit coupling within the WIEN2K code indicated that there were multiple orbital states with very small energy differences compared to room temperature; exactly which state was lowest in energy depended upon the direction chosen for the magnetization. This indicated that the system probably has disordered local spin directions and formally should be considered via a statistical average over these. In the spirit of DFT as a mean-field model, it is reasonable to average over these which, to first order, correlates to ignoring the spin-orbit contribution.

In conclusion, we combined experimental and theoretical techniques to study the valence bands of three lanthanide scandates. Our XPS and UPS spectra indicated substantial contributions from Ln4f states and O2p states, respectively. To further understand the spectra, we carried out simulations with on-site hybrid and GGA+U methods—the on-site hybrid method allowed us to accurately determine the ground state electronic structure, while the GGA+U method provided a means to account for valence band holes left during photoelectron emission. The simulated spectra matched the experimental XPS spectra well and supported the interpretation that the valence bands in $GdScO_3$, $TbScO_3$, and $DyScO_3$ consist of delocalized O2p states and Ln4f states, with small Sc3d contributions. Our findings also place minority Ln4f states close to, and in some cases at, the VBM, which can have significant implications for lanthanide scandate properties.

# 5  Acknowledgements

This work was supported by the U.S. Department of Energy, Office of Science, Basic Energy Sciences, under Award # DE-FG02-01ER45945.